\documentclass[prl,twocolumn,showpacs,preprintnumbers,amsmath,amssymb]{revtex4}

\usepackage{graphicx}
\usepackage{amsmath}
\usepackage{amsbsy}
\usepackage{dcolumn}
\usepackage{bm}

\newcommand{\BSCCO}{{Bi$_2$Sr$_2$CaCu$_2$O$_{8+x}$ }}

\newcommand{\YBCO}{{YBa$_2$Cu$_3$O$_{6+\delta}$}}

\newcommand{\LSCO}{{La$_{2-x}$Sr$_x$CuO$_4$}}

\begin{document}

\title{Disorder-Induced Static Antiferromagnetism in Cuprate Superconductors}

\author{Brian M. Andersen$^{1}$, P. J. Hirschfeld$^1$, Arno P. Kampf$^2$, and Markus Schmid$^2$}

\affiliation{$^1$Department of Physics, University of Florida,
Gainesville, Florida 32611-8440, USA\\
$^2$Theoretische Physik III, Elektronische Korrelationen und
Magnetismus, Institut f\"ur Physik, Universit\"at Augsburg, 86135
Augsburg, Germany}

\date{\today}

\begin{abstract}

Using model calculations of a disordered $d$-wave superconductor
with on-site Hubbard repulsion, we show how dopant disorder can
stabilize novel states with antiferromagnetic order. We find that
the critical strength of correlations or impurity potential
necessary to create an ordered magnetic state in the presence of
finite disorder is reduced compared to that required to create a
single isolated magnetic droplet. This may explain why in cuprates
like \LSCO (LSCO) low-energy probes have identified a static
magnetic component which persists well into the superconducting
state, whereas in cleaner systems like \YBCO (YBCO) it is absent or
minimal.

\end{abstract}


\maketitle

The occurrence of  high-temperature superconductivity in cuprates
near the antiferromagnetic (AF) phase of the parent compounds has
prompted speculation since their discovery that superconductivity
and magnetism were intimately related.  For the most part, it has
been assumed that the two forms of order compete and do not coexist,
consistent with the vanishing of the N\'eel temperature $T_N$ before
the onset of a superconducting critical temperature $T_c$, and the
suppression of $T_c$ near doping $x=1/8$ where static stripe phases
can be stable\cite{JTranquada:1995}. On the other hand, there have
been persistent reports of static AF at low temperatures $T$ in the
superconducting phase at low doping, as measured by muon spin
resonance ($\mu$SR)\cite{CPanagopoulos:2002,CPanagopoulos:2005},
nuclear magnetic resonance
(NMR)\cite{VFMitrovic:2003,KKakuyanagi:2003}, and elastic neutron
scattering (NS)\cite{BLake:2002,SWakimoto:2001}. The NS experiments
reveal an incommensurate (IC) ordering wavevector evident by a
quartet of peaks surrounding $(\pi,\pi)$. Since the neutron response
is enhanced by an applied magnetic
field\cite{BLake:2002,BLake:2001,khaykovich:2002} several authors
have discussed it in terms of coexisting $d$-wave superconductivity
(dSC) and field-induced spin density waves\cite{EDemler:2001}.
Recent magnetic Raman scattering data on LSCO has been discussed in
terms of such effects as well\cite{LHMachtoub:2005}. However, static
order also exists at zero field in the underdoped phase of
LSCO\cite{BLake:2002,SWakimoto:2001}, and has been attributed to
disorder\cite{BLake:2002}. In optimally doped LSCO, Kimura {\sl et
al.}\cite{HKimura:2003} did not detect ordered moments in pure and
1\% Zn-substituted samples.  An elastic peak similar to the pure
underdoped material was observed when 1.7\% Zn was added,
however\cite{HKimura:2003}.

Phenomena similar to those in LSCO have been observed in other
materials, e.g. Y$_{1-x}$Ca$_x$Ba$_2$Cu$_3$O$_6$, and \BSCCO (BSCCO)
where $\mu$SR directly reveals a slowing down and subsequent
freezing of spin fluctuations as $T$ is
lowered\cite{CPanagopoulos:2002,ChNiedermayer:1998}. On the other
hand, experiments on optimally doped YBCO, even with significant
percentages of Zn, have never detected static magnetic signals.
While there have been reports of AF coexisting with dSC in
underdoped YBCO, recent NS measurements on YBCO$_{6.5}$ found that
AF order, while static from the point of view of NS timescales,
$10^{-10}$s, was fluctuating faster than the timescale, $10^{-6}$s,
for $\mu$SR. Thus, it appears that while in YBCO low-frequency AF
fluctuations are present, they do not ``freeze out". Recently,
reports of static magnetism in this system near O content 6.35,
close to the onset of superconductivity, have been reported by
$\mu$SR\cite{SSanna:2004,RIMiller:2006}, but are still
controversial; the main point is that the spin-glass phase is
minimal in YBCO compared to LSCO\cite{JBobroff:2001}. 
There are many differences between YBCO and the other cuprates, of
course, but most importantly YBCO appears to be the cleanest
material because the O dopants can order in the CuO
chains. LSCO, on the other hand, is doped by randomly located
charged Sr ions only 2.4 \AA~ away from the CuO$_2$ planes. Therefore, it seems
likely that disorder itself may be responsible for inducing the
magnetism in zero field in LSCO, and possibly part of the magnetic
field dependence as well.

$\mu $SR experiments\cite{CPanagopoulos:2005} indicate that several
changes occur with decreasing $T$ for a given sample: one, where
random freezing of moments occur, and a second, where a sharply
peaked magnetization distribution arises. We propose that the first
is due to the creation of isolated AF droplets, and the second to
the formation of networks of such states which exhibit quasi-long
range order. The latter is a subtle process, which, even for a
single pair of impurities, depends on the relative orientation and
distance of impurity positions\cite{YChen:2004}. However, the basic
physics of ordering is easy to understand.  In one dimension Shender
and Kivelson\cite{EFShender:1991} pointed out that the interactions
between impurities in a quantum spin chain are non-frustrating: if
an impurity creates a local AF droplet, a second one can always
orient itself to avoid losing exchange energy. In two dimensions
(2D) this continues to apply for spin models with nearest neighbor
exchange, but may break down in the presence of mobile charges.

Below we study the disorder-induced magnetization in a dSC with
correlations described by the Hubbard model treated in an
unrestricted Hartree-Fock approximation. 
We focus on important qualitative differences between the clean and
dirty limits of both the underdoped and optimally doped regimes.
Specifically, we show that with a fixed choice of realistic
parameters, as in LSCO, the magnetism is present at low doping,
disappears at optimal doping, but can be recreated with a small
concentration of strong scatterers. For YBCO such effects are absent
since the disorder potential in the CuO$_2$ planes is negligible and
independent of doping. This result agrees with a recently proposed
origin of the unusual transport measurements in LSCO compared to
YBCO\cite{BMAndersen:2006,XSun:2006}.

\begin{figure}[b]
\begin{center}
\leavevmode
\begin{minipage}{.28\columnwidth}
\includegraphics[clip=true,width=.99\columnwidth]{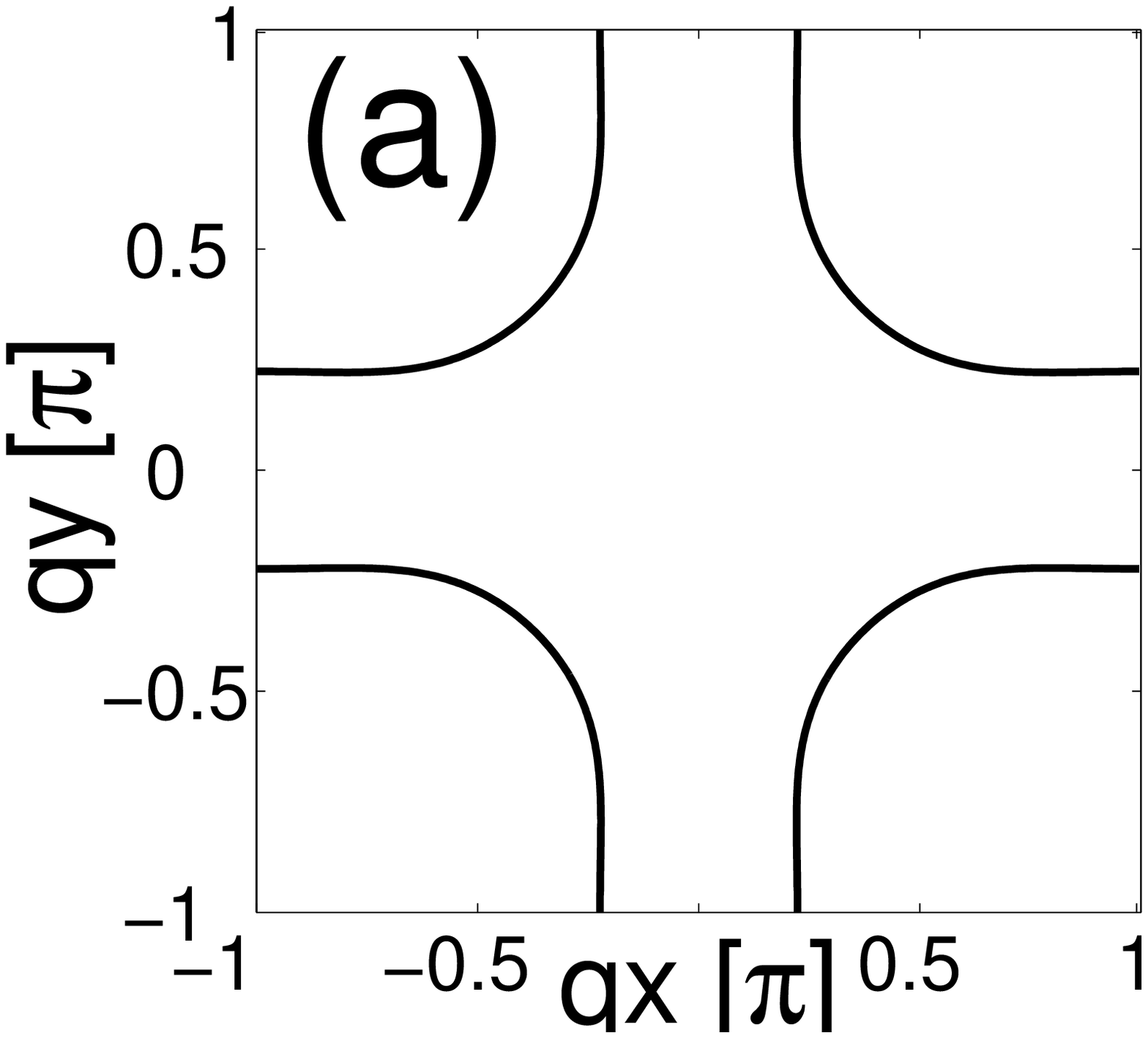}
\end{minipage}
\begin{minipage}{.33\columnwidth}
\includegraphics[clip=true,width=.99\columnwidth]{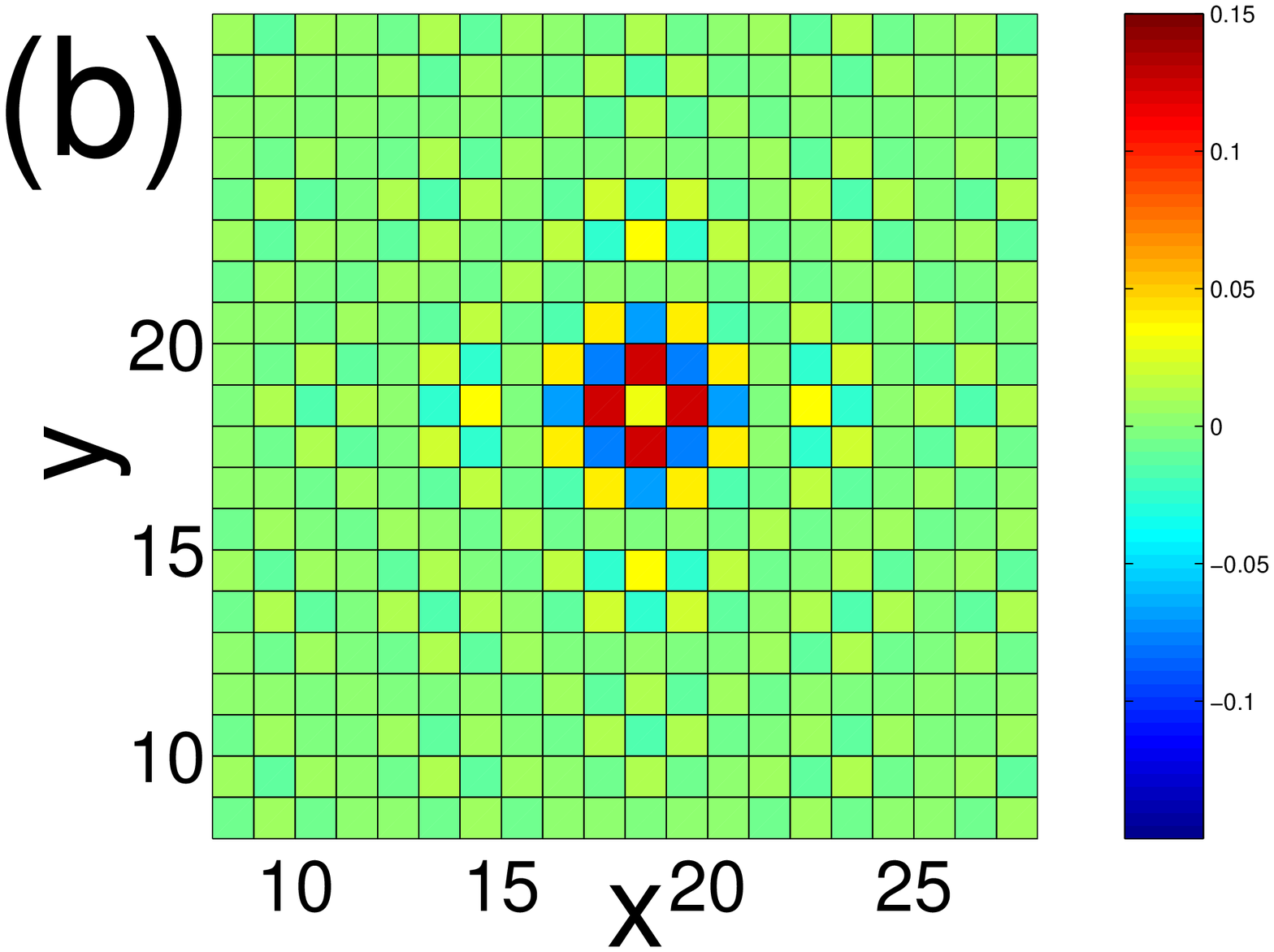}
\end{minipage}
\begin{minipage}{.33\columnwidth}
\includegraphics[clip=true,width=.99\columnwidth]{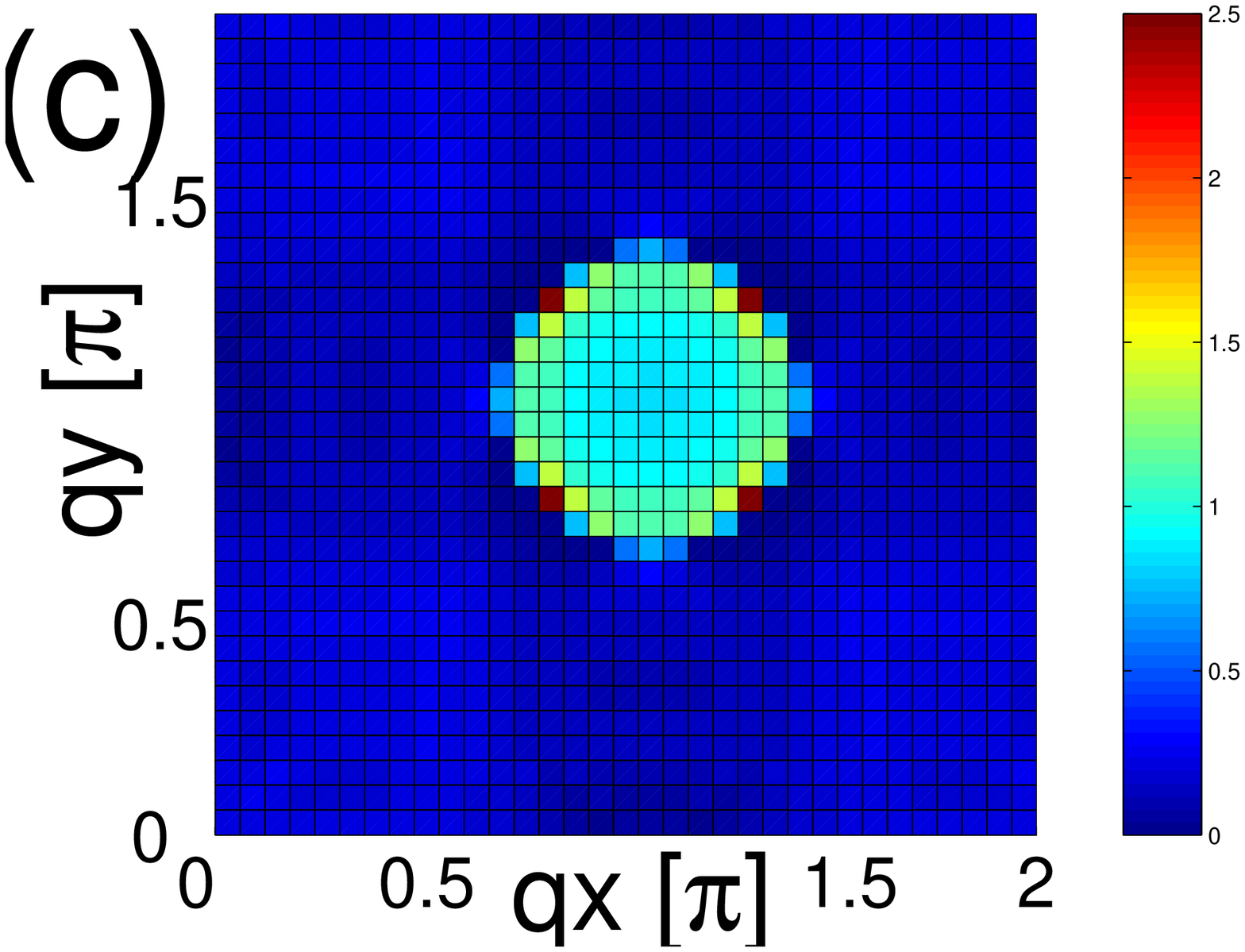}
\end{minipage}
\caption{(Color online) (a) Fermi surface with $x=7.5\%$ for the band used in this
paper. For this band the critical value for bulk order is $U_{c2}=3.50t$. (b,c) Magnetization and structure factor
$S(q)$ for a single point-like scatterer $V^{imp}=3.0t$
for $x=7.5\%$ and $U=3.3t$. For these parameters, $U_{c1}=3.25t$.}\label{fig:intro}
\end{center}
\end{figure}

{\it Model.} The Hamiltonian, defined on a 2D lattice, is
\begin{eqnarray}\label{Hamiltonian}
\hat H = &-&\sum_{\langle ij \rangle \sigma}t_{ij}\hat
c_{i\sigma}^\dagger\hat c_{j\sigma} + \sum_{i\sigma} \left(
Un_{i,{-\sigma}} + V^{imp}_i - \mu\right)\hat
c_{i\sigma}^\dagger\hat c_{i\sigma}
\nonumber\\
&+& \sum_{\langle ij\rangle}\left( \Delta_{ij}\hat
c_{i\uparrow}^\dagger \hat c_{j\downarrow}^\dagger +
\mbox{H.c.}\right).
\end{eqnarray}
Here, $\hat{c}_{i\sigma}^\dagger$ creates an electron on site $i$
with spin $\sigma$, and $t_{ij}=\{t,t',t''\}$ denote the three
nearest neighbor hopping integrals, $V^{imp}_i=\sum_{j=1}^N V^{imp}
\delta_{ij}$ is a nonmagnetic impurity potential resulting from a
set of $N$ point-like scatterers of strength $V^{imp}$, $\mu$ is the
chemical potential adjusted to fix the doping $x$, and $\Delta_{ij}$
is the $d$-wave pairing potential between sites $i$ and $j$. The
amplitude of $\Delta_{ij}$ is set by the dSC coupling constant
$V_d$\cite{JWHarter:2006}. We fix the band $t'=-0.4t$ and
$t''=0.12t$, giving the Fermi surface shown in Fig.
\ref{fig:intro}(a). We have solved Eq.(\ref{Hamiltonian})
self-consistently by diagonalizing the associated Bogoliubov-de
Gennes equations in the $T=0$ limit on $34\times 34$
systems\cite{JWHarter:2006}.

Eq.(\ref{Hamiltonian}) has been used extensively to study bulk
competing phases, field-induced magnetization, as well as novel
bound states at interfaces between AF and
superconductors\cite{allHamiltonian}. It has also been used to study
field-induced moment formation around nonmagnetic impurities in the
correlated dSC\cite{YOhashi:2002,YChen:2004}, with considerable
success at fitting NMR lineshapes\cite{JWHarter:2006} in Li- and
Zn-substituted YBCO. Interference effects of these paramagnetic
states, while significant, are less important than in the heavily
disordered case studied in this paper where the ground state in zero
field exhibits local magnetism.

{\it Results.} For clean systems it is well-known that there exists
a large region in $(U,V_d,T)$-space dominated by spin and charge ordered
stripe states of coexisting dSC and AF order\cite{allHamiltonian}.
For fixed $V_d$ and $T$ we denote by $U_{c2}$ the critical Coulomb
repulsion for entering this bulk magnetic state, i.e. for
$U<U_{c2}$, in the absence of disorder, the ground state is a
homogeneous dSC.

For a single nonmagnetic impurity\cite{Balatskyreview}, there
exists a lower critical $U_{c1}$ such that local impurity-induced
magnetization exists for $U_{c1}<U<U_{c2}$\cite{ZWang:2002} as
shown in Fig. \ref{fig:intro}(b). In general $U_{c1}$ depends on
the impurity strength $V^{imp}$\cite{YChen:2004,JWHarter:2006}.
However, already for two impurities interference effects modify
the 1-impurity phase diagram\cite{YChen:2004}. The magnetic
structure factor $S(q)$ associated with the AF droplet in Fig.
\ref{fig:intro}(b) is shown in Fig. \ref{fig:intro}(c). It is
dominated by four IC peaks along the diagonals, a result which is,
however, sensitive to the input parameters.

We now turn to the many-impurity situation. In the following model
for LSCO the Sr ions are assumed to be the primary source of
disorder, such that $n^{imp}=x$, where
$n^{imp}$ denotes the impurity concentration.
\begin{figure}[t]
\begin{minipage}{.49\columnwidth}
\includegraphics[clip=true,width=.92\columnwidth]{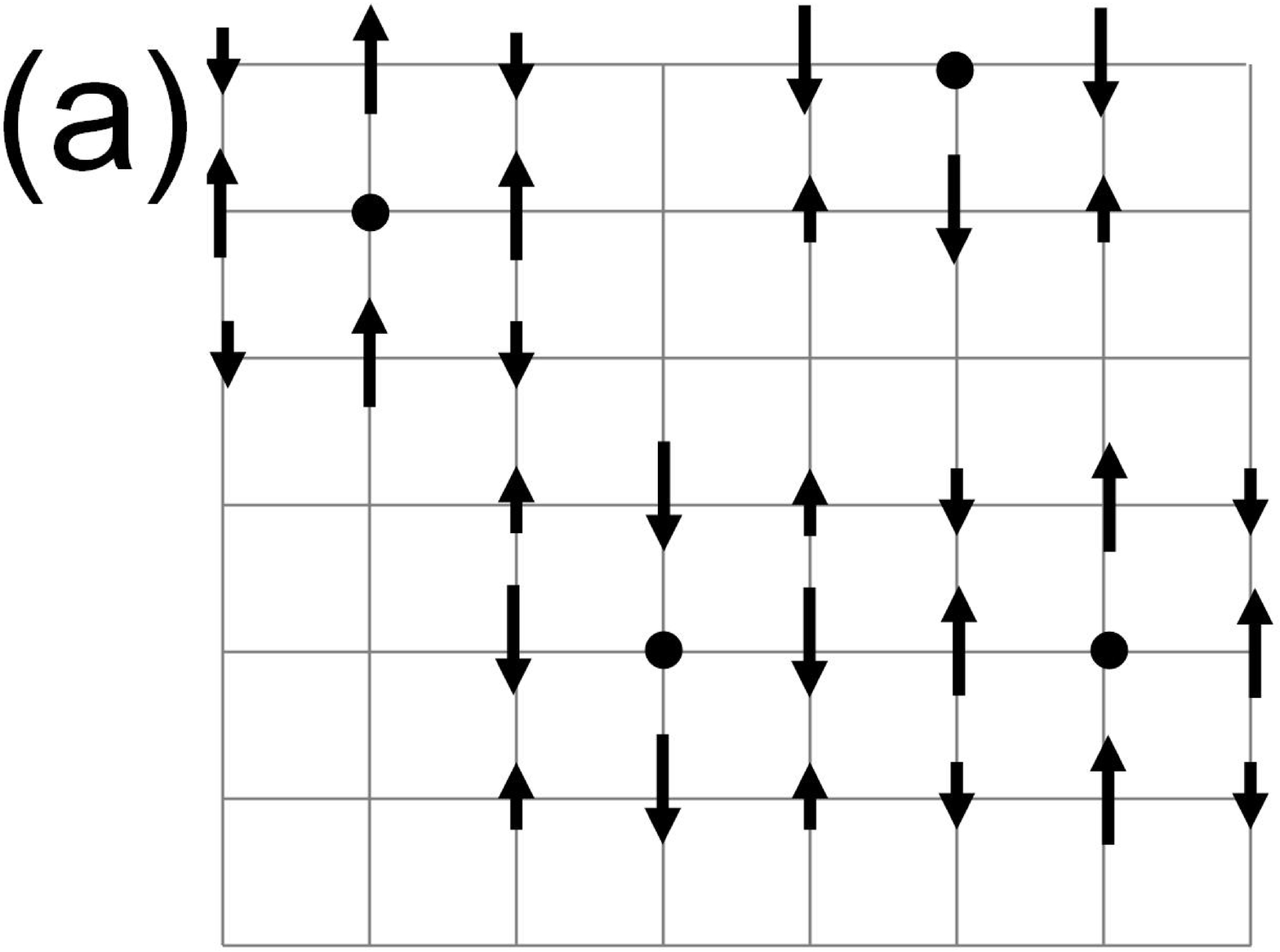}
\end{minipage}
\begin{minipage}{.49\columnwidth}
\includegraphics[clip=true,width=.99\columnwidth]{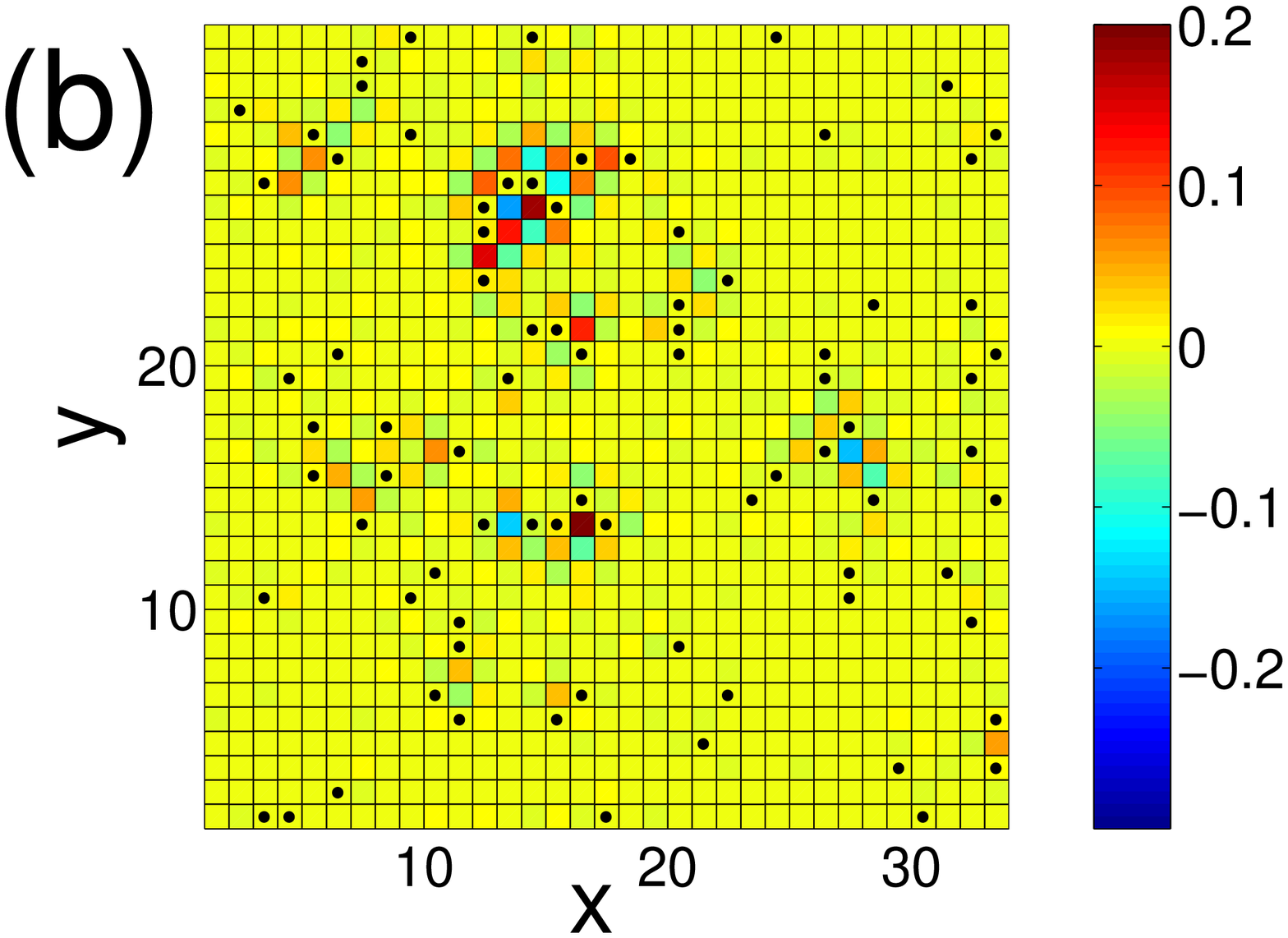}
\end{minipage}\\
\begin{minipage}{.49\columnwidth}
\includegraphics[clip=true,width=.99\columnwidth]{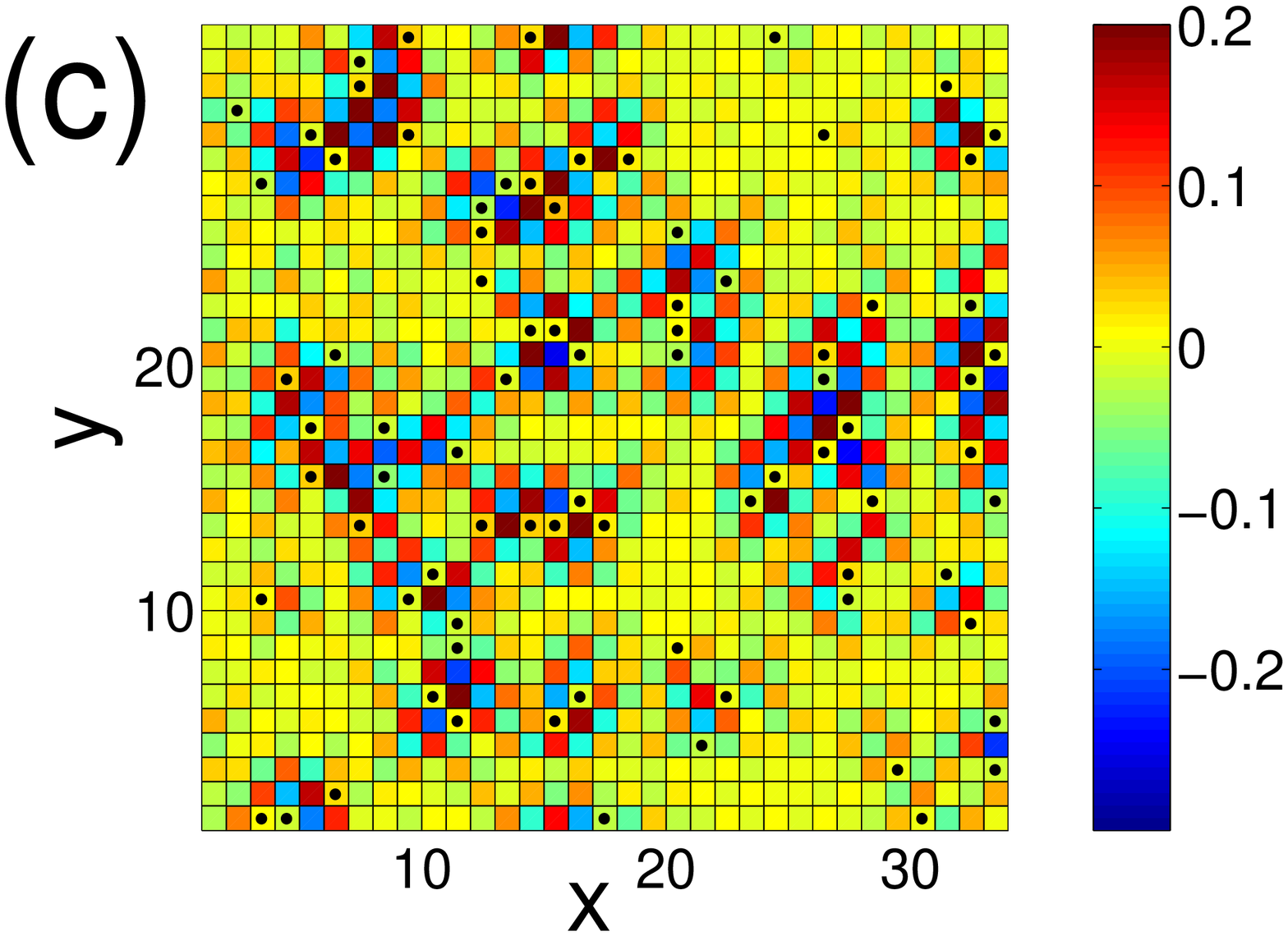}
\end{minipage}
\begin{minipage}{.49\columnwidth}
\includegraphics[clip=true,width=.99\columnwidth]{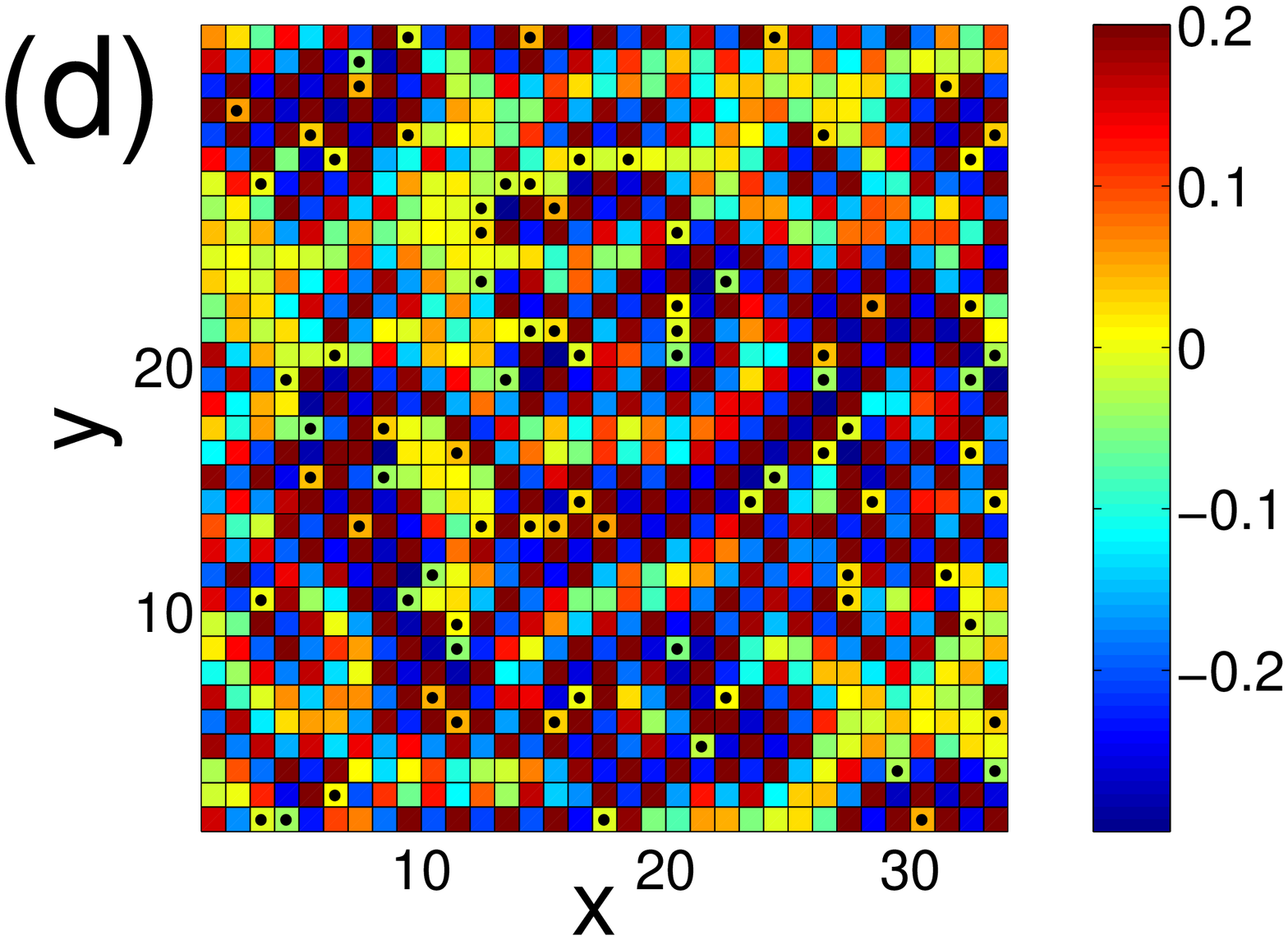}
\end{minipage}
\caption{(Color online) (a) Schematic: stabilization of single
N\'eel phase by impurities. (b-d) Disorder-induced magnetization
for a single fixed impurity configuration with $x=7.5\%$ and
$U=2.8t$ (b), $U=3.2t$ (c), and $U=3.6t$ (d).  In (a-d) the
black dots indicate the impurity/dopant positions.}
\label{fig:MagnvsU}
\end{figure}
These systems are in the strongly disordered regime where the Fermi
wavelength $\lambda_F$ is comparable to the average distance between
the dopants, such that the disorder is far from the 1-impurity limit
and novel emergent spin-glass states
can arise. 
Since the Sr dopants are removed from (but close to) the CuO$_2$
planes we model them as intermediate strength scatterers with
$V^{imp}=3.0t$ (below we use $V^{dop}$ to indicate the Sr potential). 
Fig. \ref{fig:MagnvsU}(a) shows the simplest schematic picture of disorder
stabilization of a single N\'eel phase\cite{EFShender:1991} by
nonmagnetic impurities. Not all impurities in the
correlated system need ``magnetize" for a given $U$, however: in
the disordered system, the effective criterion to drive the
impurity through the local magnetic phase transition is different
for each impurity. Increasing the repulsion $U$ then increases the
concentration of impurities which induce a local magnetization
droplet, as shown in Fig. \ref{fig:MagnvsU}(b,c). With further
increase of $U$, the system evolves from a state with dilute
non-overlapping AF droplets to connected spin textures (Fig.
\ref{fig:MagnvsU}(d)). The resulting patterns exhibit AF domain
structure and are  more complex than suggested in Fig. \ref{fig:MagnvsU}(a), due to
frustration induced by the charge degrees of freedom
in Eq.(\ref{Hamiltonian}), and/or glassy supercooling effects.

While the current mean field treatment of the Hubbard model does not
faithfully capture the band narrowing due to correlations near
half-filling which leads to the Mott transition, it may be expected
that underdoped systems are characterized by larger effective
interactions.  In our picture for LSCO, the $x$ dependence of the
spin order is therefore described qualitatively by the sequence
\ref{fig:MagnvsU}(d) $\rightarrow$ \ref{fig:MagnvsU}(c)
$\rightarrow$ \ref{fig:MagnvsU}(b), until it disappears completely
at effective $U$'s below $U_{c1}$ near optimal doping. Increasing
$x$ should also be accompanied by a weakening of the Sr potential
$V^{dop}$ due to enhanced screening. Within our model, increasing
$U$ or $V^{dop}$ leads to qualitatively similar results, and we
cannot determine from this approach which effect is dominant in real
systems. Note from Fig. \ref{fig:MagnvsU} that AF droplets are
induced for $U\gtrsim 2.4t$, a substantially reduced critical value
compared to the 1-impurity case in Fig. \ref{fig:intro}. This is
because the Hubbard correlations induce charge redistributions which
alter the effective local chemical potential, such that the
criterion for the magnetization of each impurity depends on its
local disorder environment. Some regions containing impurities have
charge densities closer to the phase boundary for AF order, thus
enhancing local moment formation relative to the single impurity
case. In the limit of large $U$,  the  magnetic order becomes
qualitatively similar to that arising in a stripe state with
quenched disorder\cite{JARobinson:2006}.

\begin{figure}[t]
\begin{center}
\leavevmode
\begin{minipage}{.49\columnwidth}
\includegraphics[clip=true,width=.99\columnwidth]{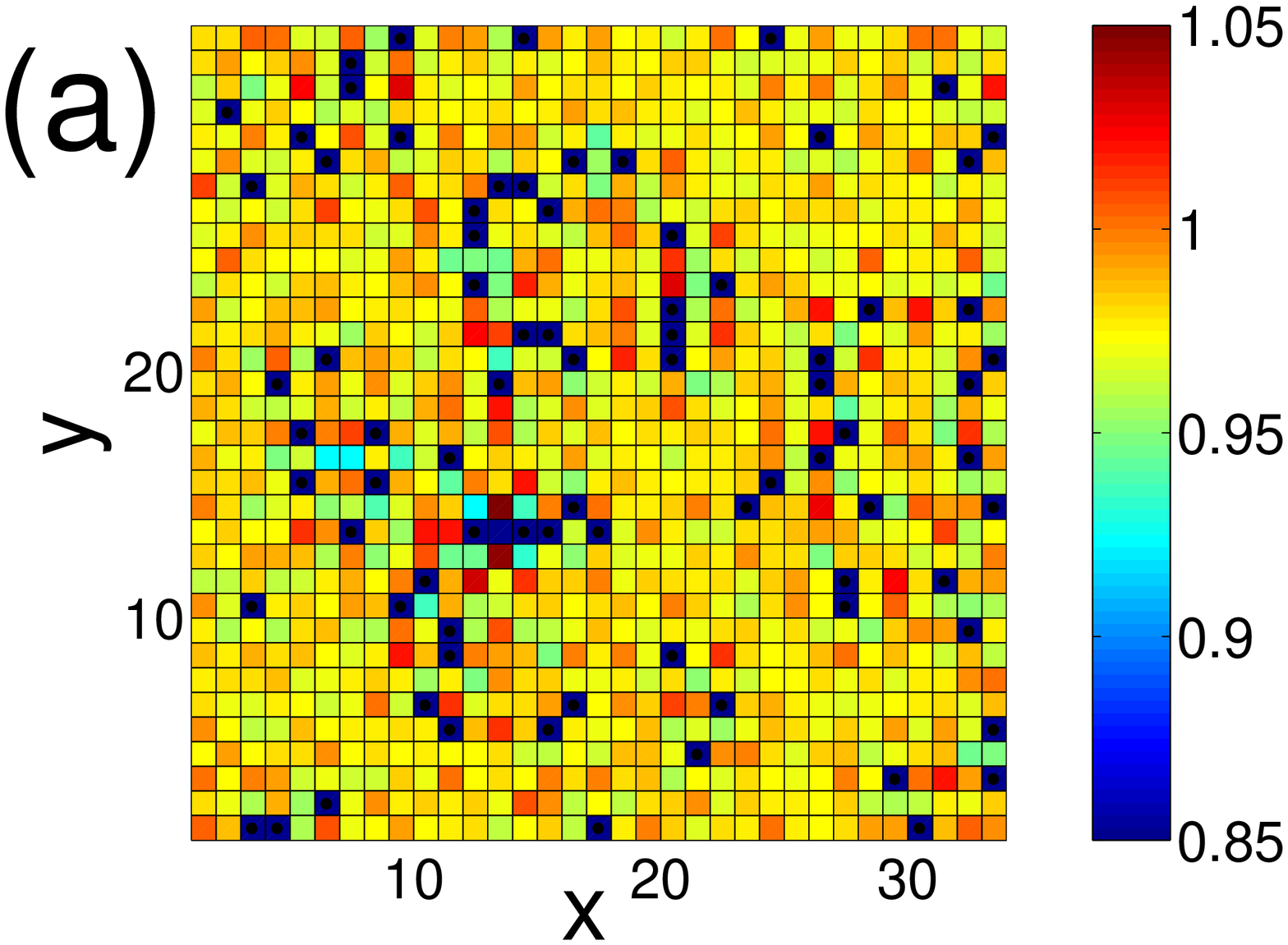}
\end{minipage}
\begin{minipage}{.49\columnwidth}
\includegraphics[clip=true,width=.99\columnwidth]{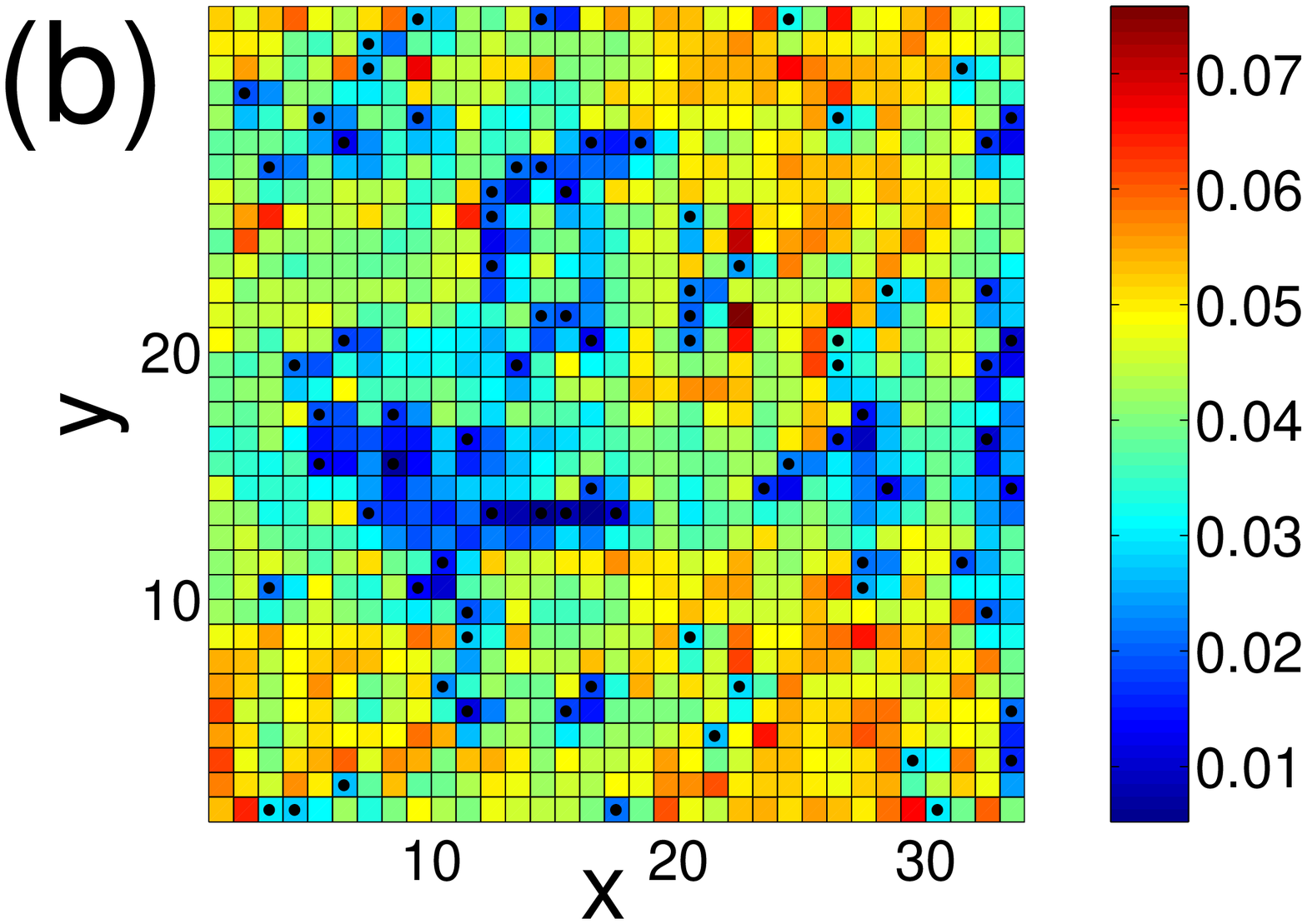}
\end{minipage}\\
\begin{minipage}{.49\columnwidth}
\includegraphics[clip=true,width=.99\columnwidth]{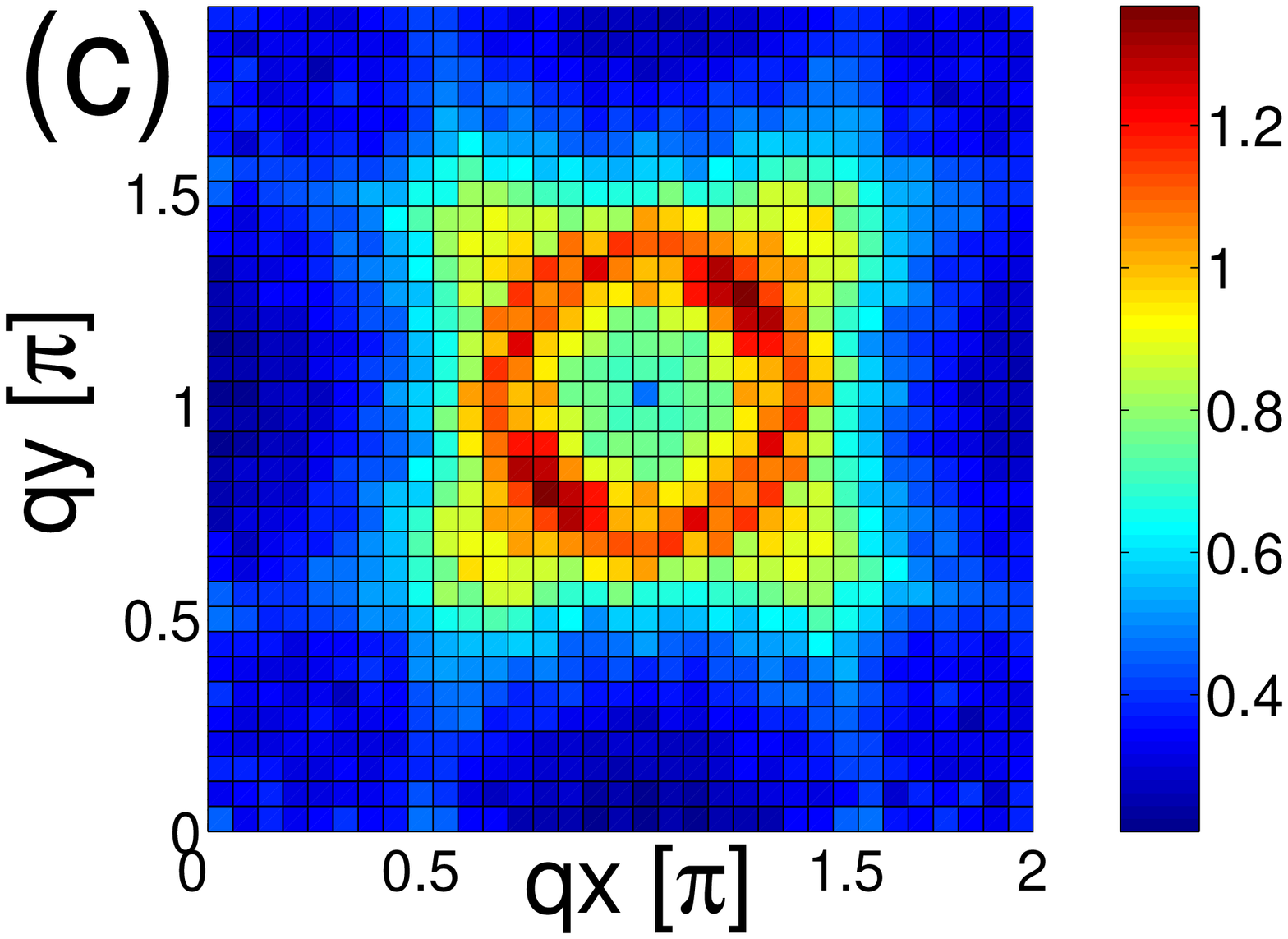}
\end{minipage}
\begin{minipage}{.49\columnwidth}
\includegraphics[clip=true,width=.99\columnwidth]{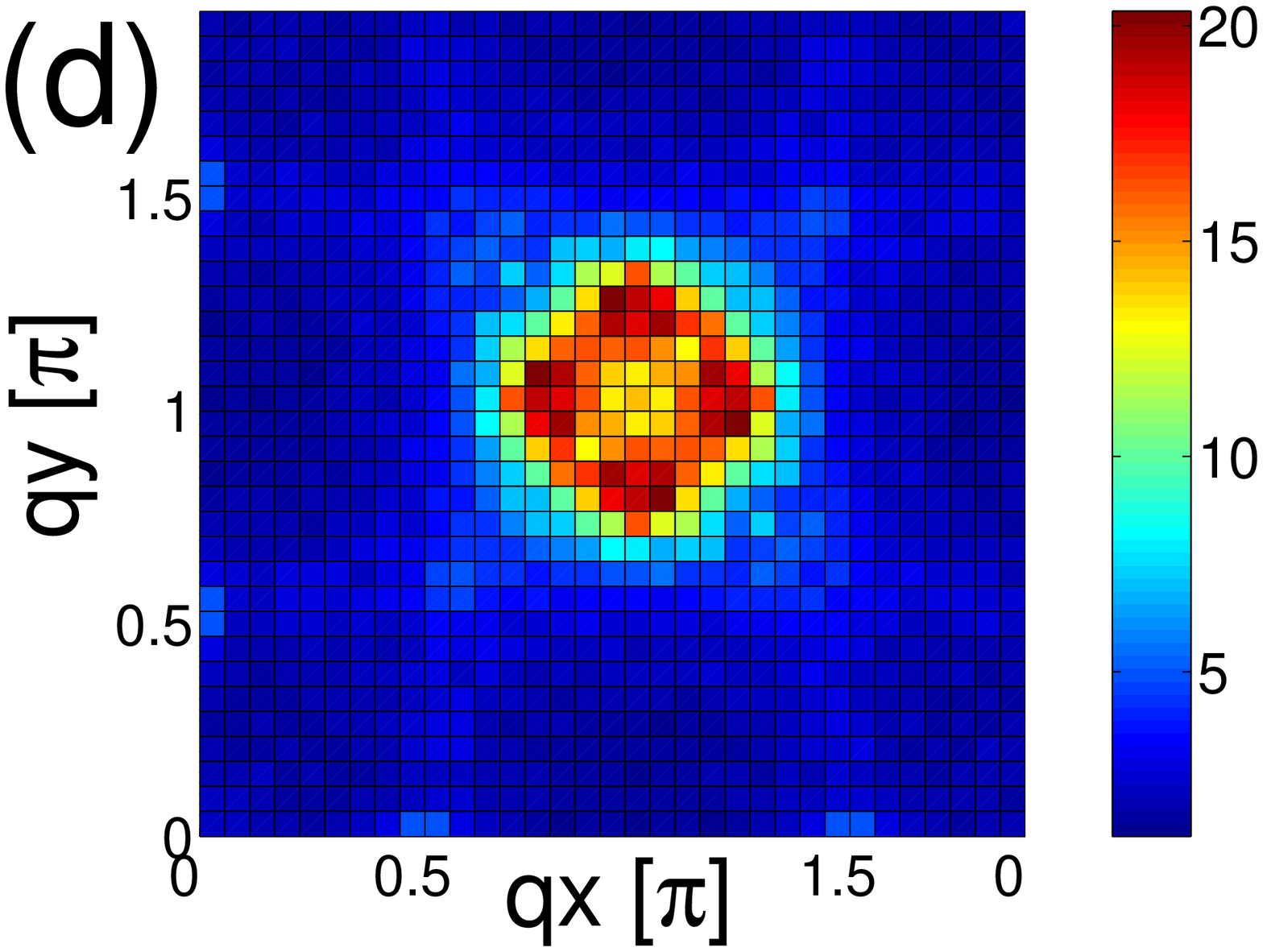}
\end{minipage}
\caption{(Color online) (a) Electronic density and (b) dSC gap for
the parameters shown in Fig. \ref{fig:MagnvsU}(c). (c,d) display
the impurity averaged magnetic structure factor $S(q)$ for the
dilute magnetic droplet limit (small $U$) and dense limit (large
$U$) corresponding to Fig. \ref{fig:MagnvsU}(b) and Fig.
\ref{fig:MagnvsU}(d), respectively.}\label{fig:elecdeltaSq}
\end{center}
\end{figure}

In Fig. \ref{fig:elecdeltaSq}(a,b) we show the inhomogeneous
electronic density and dSC gap corresponding to Fig.
\ref{fig:MagnvsU}(c). The repulsive Sr dopants locally suppress
both quantities which are therefore anti-correlated with the
induced spin order. As expected, the gap varies more smoothly
compared to the density due to the coherence length $\xi$ of the
dSC condensate. The magnetic structure factor $S(q)$ (averaged
over 30 distinct impurity configurations) associated with the
disorder induced magnetic order in Fig. \ref{fig:MagnvsU}(b,d) is
shown in Fig. \ref{fig:elecdeltaSq}(c,d), respectively. It is
dominated by an IC ring surrounding $(\pi,\pi)$ with an intensity
distribution similar to the single impurity $S(q)$ for the dilute
case (Fig. \ref{fig:elecdeltaSq}(c)), but which rotates into a $+$
shaped pattern for the connected spin textures at larger $U$ (Fig.
\ref{fig:elecdeltaSq}(d)). This implies a rotation (and weakening)
of the IC pattern close to the region where the static order
disappears. As mentioned above, such details are, however,
sensitive to e.g. the specific band parameters\cite{footnote1}.

\begin{figure}[t]
\includegraphics[clip=true,width=.95\columnwidth]{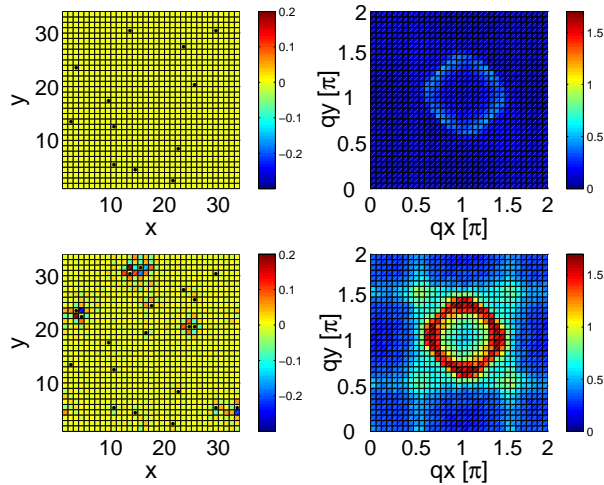}
\caption{(Color online) Induced magnetization (left) and impurity
averaged $S(q)$ (right) for $U=3.2t$ with $1\%$ (top) and $2\%$
(bottom) concentration of randomly distributed strong scatterers.
Note that the figures are shown on the same scale. For clarity, the
black dots (left) only show the Zn positions.}
\label{fig:Magnvsconc1}
\end{figure}

\begin{figure}[t]
\includegraphics[clip=true,width=.95\columnwidth]{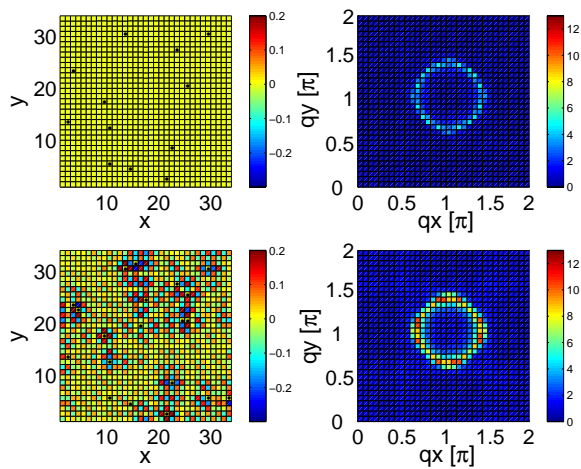}
\caption{(Color online) Same as Fig. \ref{fig:Magnvsconc1} but for
$U=3.4t$.} \label{fig:Magnvsconc2}
\end{figure}

Lastly we discuss optimally doped LSCO where static
AF is absent in nominally clean samples, but where it can be
induced by magnetic fields\cite{BLake:2002} or critical
concentrations of strong scatterers. For instance, Kimura {\sl et
al.}\cite{HKimura:2003} found that for $x=0.15$ it takes
approximately $2\%$ Zn to induce IC peaks in the NS
diffraction. Below this critical concentration there was no
measurable signal above the background. We have simulated this
situation by solving Eq.(\ref{Hamiltonian}) with
$x=0.15$ in the presence of $1\%$ and $2\%$ randomly distributed
strong scatterers ($V^{Zn}=100t$) in addition to the
$n^{imp}=15\%$ weak ($V^{dop}=1t$) dopant
impurities. As mentioned above the weaker $V^{dop}$ compared to the
underdoped case is expected from an enhanced screening of the Sr
potential at optimal doping. For a single Zn impurity, magnetization
is induced only for $U\gtrsim 3.35t$. The many-impurity results
including both $V^{dop}$ and $V^{Zn}$ are shown for different
Coulomb repulsions in Fig. \ref{fig:Magnvsconc1} and Fig.
\ref{fig:Magnvsconc2}. For systems with $1\%$ Zn Fig.
\ref{fig:Magnvsconc1}(a-b) reveals a negligible induced
magnetization. However, as seen from Fig.
\ref{fig:Magnvsconc1}(c-d), simply
adding enough strong scatterers can induce sizable local magnetic
order.
It is remarkable that such a small difference in the concentration
can induce magnetism similar to the experimental observations.
The number of Zn ions that induce AF droplets depends on the Hubbard $U$ as
seen by comparing Fig. \ref{fig:Magnvsconc1} and Fig.
\ref{fig:Magnvsconc2}. The resulting disorder-averaged $S(q)$ agrees
well with the NS measurements in Ref. \onlinecite{HKimura:2003}.

{\it Conclusions.} The interplay of dopant disorder and electronic
correlations can induce novel magnetic states which we propose
exist in intrinsically disordered cuprates like LSCO and BSCCO, in
contrast to the cleaner YBCO system.
Lastly, we showed that small concentrations of Zn can induce a
similar magnetic state which has apparently been observed in NS
experiments on optimally doped LSCO. An obvious question is to
what extent the magnetic state influences the scattering of
quasiparticles in the dSC and normal states. Studies along these
lines are in progress.

We acknowledge useful discussions with I. Affleck, M.
Gabay, P. Hedeg\aa rd, B. Keimer, T. Kopp, C. Panagopoulos, and N.
Trivedi. P.J.H. and B.M.A. were supported in part by DOE Grant
DE-FG02-05ER46236. A.P.K. and M.S. acknowledge support through
SFB 484 of the DFG.

\end{document}